\documentclass[12pt]{article}
\usepackage{epsf}
\usepackage{graphicx}
\usepackage{color}
\def\ds{\displaystyle}
\def\bea{\begin{eqnarray}}
\def\eea{\end{eqnarray}}
\def\shat{\hat{s}}

\def\mlhat{\hat{m}_\ell}
\def\plep{p^-}

\def\nnb{\nonumber}
\def\lla{\left<}
\def\rra{\right>}
\def\rar{\rightarrow}
\def\nnb{\nonumber}
\def\la{\langle}
\def\ra{\rangle}
\def\g5{\gamma_5}

\def\es{\!\!\! &=& \!\!\!}
\def\ap{\!\!\! &\approx& \!\!\!}

\def\ga{\gamma}
\def\UP{\cal U}
\def\la{\langle}
\def\ra{\rangle}
\title{ {\bf CP-Conserving Unparticle Phase Effects on the Unpolarized and Polarized Direct
CP Asymmetry in $b \rightarrow d \ell^+\ell^-$ Transition}}
\author{\vspace{1cm}\\
        {\small V. Bashiry$^1$\thanks {e-mail:
        bashiry@ciu.edu.tr}}
        \\ {\small $^1$ Engineering Faculty, Cyprus International
University,} \\ {\small Via Mersin 10, Turkey } }

\date{}
\begin{document}
\setlength{\baselineskip}{24pt}
\maketitle
\setlength{\baselineskip}{7mm}
\begin{abstract}
We examine the unparticle CP-conserving phase effects on the
direct CP asymmetry for both polarized and unpolarized lepton in
the inclusive $b\rightarrow d \ell^+ \ell^-$ transition, where the
flavor changing neutral currents are forbidden at tree level but
are induced by one-loop penguin diagrams. The averaged polarized
and unpolarized CP asymmetries depict strong dependency on the
unparticle parameters. In particular, a sizable discrepancy
corresponding to the standard model is achieved when the scale
dimension  value is $1<d_{\UP}< 2$.  We see that the unparticle
stuff significantly enhances, suppresses or changes the sign of
the CP asymmetry depending on the definite value of the scaling
dimension $d_{\UP}$. Especially, when $d_{\UP}\sim 1.1$ the CP
asymmetries vanish.

\end{abstract}
\thispagestyle{empty}
\newpage
\setcounter{page}{1}
\section{Introduction}
Georgi \cite{Georgi1,Georgi2} has recently proposed that
unparticles stuff which can couple to the standard model (SM)
particles at the Tev Scale. Unparticles are massless and invisible
coming out of a scale invariant sector with non-integer scaling
dimension $d_{\UP}$ when decoupled at a large scale. The
propagator of these invisible unparticles includes  CP conserving
phase which is dependent on the non-integer scaling dimension
$d_{\UP}$ \cite{Georgi2}. The virtual unparticle propagation and
its effects were first studied by Georgi himself\cite{Georgi2}.
Moreover, the CP conserving phase of the unparticles and its
effects in FCNC, especially, in hadronic and semileptonic B decays
 have been studied in \cite{Chen1, Chen2, Chen3, Chen4}.

A phenomenological study needs construction of the effective
Hamiltonian to describe the interactions of unparticles with the
SM fields in the low energy level\cite{Iltan:2007cb}. So that, we
can investigate the effects of the possible scale invariant
sector, experimentally.

The direct search  of the unparticles is based on the study of
missing energies at various processes which can be measured at LHC
or ILC. The indirect search includes the dipole moments of
fundamental particles, lepton flavor violation(LFV) and flavor
changing neutral current(FCNC)  processes where the virtual
unparticles enter as mediator. Note that, the phenomenological
studies considering the direct and indirect search on unparticles
have been progressing \cite{Georgi2}-\cite{unparticlenew}; their
effects on the missing energy of many processes; the anomalous
magnetic moments; the electric dipole moments; $D^0-\bar{D}^0$ and
$B^0-\bar{B}^0$ mixing; lepton flavor violating interactions;
direct CP violation in particle physics; the phenomenological
implications in cosmology and in astrophysics.

It is well known that in a decay process the existence of direct
CPA( $A_{CP}$) requires:  firstly at least two different terms in
decay amplitude. Secondly, These terms must depend on two types of
phases named weak($\delta$) and strong($\phi$) phases. The weak
phase is CP violating and strong phase is CP conserving phase. The
$A_{CP}$ depends on the interference of different amplitude and is
proportional to the phases. i.e.,
\begin{equation}\label{acp}
   A_{CP}\propto \sin(\delta)\sin(\phi)
\end{equation}
The sizable value of $A_{CP}$ can be obtained if both phases are
non-zero and large. The weak phase of the SM is a unique phase of
the Cabibbo-Kobayashi-Maskawa(CKM) quark mixing matrix. This weak
phase is a free parameter of the SM and can not  be fixed by SM
itself but it has been fixed by experimental
methods\cite{PDG2006}. Unlike the weak Phase, the CP conserving
strong phase is process dependent(not unique). The theoretical
calculation of the strong phase is in  general hard due to the
hadronic uncertainty. The CP conserving unparticle phase exist in
the propagators beside the strong phase can affect the value of
the $A_{CP}$ in some decay processes(see Eq. \ref{acp}). To
explore this possibility, Chuan-Hung Chen, {\it et al.}
concentrated on some pure hadronic and pure leptonic B
decays\cite{Chen1, Chen5}.

We aim to study the possible effects of the CP conserving phase in
semileptonic B decays.
 Rare semileptonic decays $b\rightarrow
 s(d)\ell^{+}\ell^{-}$ are more informative for this aim, since
these decays are relatively clean compared to pure hadronic
decays. It is well known that the matrix element for the
$b\rightarrow s\ell^{+}\ell^{-}$ transition involves only one
independent CKM matrix element, namely $|V_{tb}V^{*}_{ts}|$, so
the CP-violation in this channel is strongly suppressed in the SM
considering the above mentioned requirements of the CPA which
requires the weak phase. However, the possibility of CP-violation
as a result of the new weak phase coming out of the physics beyond
the standard model in $b\rightarrow s$ transition has been studied
in supersymmetry\cite{Kruger:2000zg}-\cite{Kruger:2000ff},
fourth-generation standard
model(SM4)\cite{Arhrib:2002md}-\cite{Zolfagharpour:2007eh} and
minimal extension of the SM\cite{Bashiry:2006wd}. Situation for
$b\rightarrow d\ell^{+}\ell^{-}$ is totaly different from
$b\rightarrow s\ell^{+}\ell^{-}$ transition. In this case, all CKM
matrix elements $|V_{td}V^{*}_{tb}|$, $|V_{cd}V^{*}_{cb}|$ and
$|V_{ud}V^{*}_{ub}|$ are in the same order and for this reason the
matrix element of $b\rightarrow d\ell^{+}\ell^{-}$ transition
contains two different amplitudes with two different CKM elements
and therefore sizable  CPA is expected\cite{babu, Aliev:2003hw}.
Here, we study the effects of the CP conserving unparticle phase
on CP asymmetry in the $b\rightarrow d\ell^{+}\ell^{-}$ transition
with unpolarized and polarized lepton cases.

This study encompasses four sections: In Section 2, we present the
the effective lagrangian and effective vertices which drive the
FCNC decays with vector unparticle mediation. In section 3, we
calculate the polarized and unpolarized CP asymmetries.  Section 4
is devoted to the discussion and our conclusions.

\section{ Flavor changing neutral currents mediated by vector unparticle } \label{sec:loop}
The starting point of the idea is the interaction between two
sectors, the SM and the ultraviolet sector with non-trivial
infrared fixed point, at high energy level. The ultraviolet sector
appears as new degrees of freedom, called unparticles, being
massless and having non integral scaling dimension $d_{\UP}$
around, $\Lambda_{\UP}\sim 1\,TeV$. This mechanism results in the
existence of the  effective field theory with effective Lagrangian
in the low energy level.  One may for simplicity  assume that
unparticles only couple to the flavor conserving fermion
currents\cite{Chen5},
described by \cite{Georgi1,Georgi2,Cheung1,Cheung2}
 \bea
   \frac{1}{\Lambda^{d_{\UP}-1}_{\UP}}\bar f \ga_{\mu} \left( C^{\rm f}_{L} P_L
   +C^{\rm f}_{R} P_R\right) f O^{\mu}_{\UP} \label{eq:lang_UP}
 \eea
where  $O^{\mu}_{\UP}$ is the unparticle operator.  Similar to the
SM, FCNCs such as $f\to f'\, \UP $ can be induced by the charged
weak currents at quantum loop level and clearly, neutral current
$f\to f\, \UP $ is flavor diagonal.

\begin{figure}[hpbt]
\begin{center}
\includegraphics*[width=2 in]{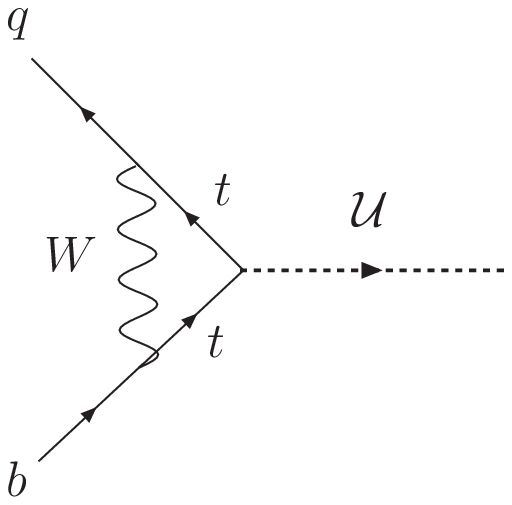}
\caption{Feynman diagram for $b\to q(s~ \texttt{or} ~d) \UP$,
where $t$ is top quark.}
 \label{fig:up}
 \end{center}
\end{figure}

%
The leading order of effective Hamiltonian for the
Fig.~\ref{fig:up}
 can be written as follows:
 \bea
{\cal L}_{\UP}= \frac{g^2}{\Lambda^{d_{\UP}-1}_{\UP}}V_{tb}
V^{*}_{tq}C^{qb} _{L} \bar{q} \ga_{\mu}  P_L b\, O^{\mu}_{\UP} \,,
 \label{eq:efflang}
 \eea
where
  \bea
  C^{qb}_{L}&=& \frac{1}{(4\pi)^2} I(x_t )  \,,\nnb\\
 I(x_t)&=&\frac{x_t(2 C^{t}_R +C^{t} _{L} x_t)}{2(1-x_t)^2} \left( -1+x_t -\ln x_t \right)\,,
 \eea
with $x_t=m^2_t/m^2_W$\cite{Chen5}.

To obtain the effective Hamiltonian for $b\to q f\, \bar f$
transition where unparticles enter as mediators, we must obtain
the unparticle propagator, which
 is given by \cite{Georgi1,Georgi2,Cheung1,Cheung2}
  \bea
  \int d^4x e^{i p\cdot x} \la 0| T\left( O^{\mu}_{\UP}(x)
O^{\nu}_{\UP}(0)\right) |0 \ra
 = i\Delta_{\UP}(p^2)\,e^{-i\phi_{\UP}}
\label{eq:uprop}
 \eea
where
 \bea\label{propage}
    \Delta_{\UP}(p^2)&=& \frac{A_{d_{\UP}}}{2\sin(d_{\UP}
\pi)} \frac{ -g^{\mu \nu} + a p^{\mu} p^{\nu}/p^2 }{(p^2
+i)^{2-d_{\UP}}}\,,
\nnb\\
\phi_{\UP}&=&(d_{\UP}-2)\pi \,,
 \eea
where $a=1$ for transverse $O^{\mu}_{\UP}$ and
$a=\frac{2(d-2)}{d-1}$ in the conformal field
theories(CFT)\cite{Grinstein}. Note also that, the contribution
from the longitudinal piece $a p^{\mu} p^{\nu}/p^2 $ in Eq.
(\ref{propage}) can be dropped for massless or light external
fermions. In this case, Georgi\cite{Georgi2} and Grinstein {\it
et. al,}\cite{Grinstein} approaches provide the same result. Also,
 \bea
A_{d_{\UP}}&=&  \frac{16 \pi^{5/2}}{(2\pi)^{2d_{\UP}}}
\frac{\Gamma(d_{\UP}+1/2)}{\Gamma(d_{\UP}-1) \Gamma(2 d_{\UP})}\,.
\label{eq:pro_up}
\eea
Note that, in Eq. (\ref{eq:uprop}) the phase factor arises from
$(-1)^{d_{\UP}-2}=e^{-i\pi(d_{\UP}-2)}$ and, here, the massless
vector unparticle operator is conserved current, i.e.,
$\partial_{\mu} O^{\mu}_{\UP}=0$. The effective Hamiltonian for
$b\to q f\, \bar f$ just with the contribution of the vector
unparticle as a mediator can be given as:
 \bea
{\cal H}_{\UP}&=& - \frac{G_F}{\sqrt{2}}V_{tb} V^{*}_{tq}
\tilde{\Delta}_{\UP}(p^2) e^{-i\phi_{\UP}}\, \bar q \ga_{\mu}
P_{L} b\; \bar f\, \ga^{\mu}\left( C^{\rm f}_{L} P_L + C^{\rm
f}_{R} P_R \right)\, f\,, \label{eq:h_U}
 \eea
where  \bea\label{propage2} \tilde{\Delta}_{\UP}(p^2)=8 C^{qb}_{L}
\frac{A_{d_{\UP}}}{2\sin d_{\UP}\pi} \frac{ m^2_{W}}{p^2}
\left(\frac{p^2}{\Lambda^2_{\UP}}\right)^{d_{\UP}-1}\,.
 \eea
Here, $f$ stands for fermions. i.e., $f$ can be neutrinos or
charged leptons or quarks.

\section{  $b\to d  \ell^+
\ell^-$ transition  in the presence of the vector unparticle as a
mediator }

\begin{figure}[hpbt]
\begin{center}
\includegraphics*[width=2.5 in]{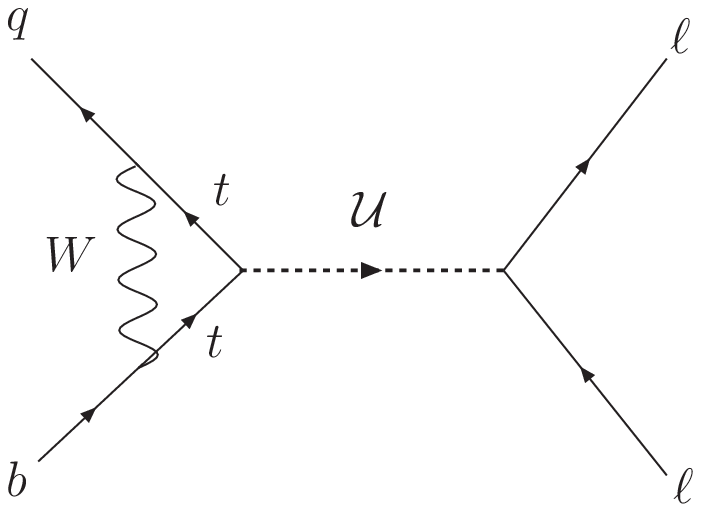}
\caption{$b\to q \ell^{+} \ell^{-}$ decays induced by unparticle
penguin diagram.}
 \label{fig:up_bqll}
 \end{center}
\end{figure}

By the extension of the  $b\to d\UP$ to study the semileptonic
decays of $b\to d \ell^+ \ell^-$, the decay  amplitude in the
presence of the vector unparticle as a mediator can be obtained.
Here, again we assume that unparticle coupled to the leptons are
flavor conserving. The penguin  diagram describing this decay is
shown in Fig.~\ref{fig:up_bqll}. Due to the CKM suppression, the
semileptonic decays with $b\to d$ are much less than those of
$b\to s$.  However, it is worth to study the $b\to d$ transition
beyond the $b\to s$ one because the CKM matrix element $V_{td}$
carries a CP violating weak phase, which is almost vanishes in the
$b\to s$ transition. Thus, $b\to d \ell^{+} \ell^{-}$ decay could
be even more interesting on CP violation in the framework of
unparticle physics. we will focus on the CP violating asymmetry in
$b\to d \ell^{+} \ell^{-}$.

%
The QCD corrected effective Lagrangian for the decays
$b\rightarrow d\ell^{+}\ell^{-}$ can be obtained by integrating
out the heavy quarks and the heavy electroweak bosons are reads as
follows in the SM:

\begin{eqnarray}
 M =\frac{G_{F}\alpha_{em}
\lambda_t}{\sqrt{2}\pi}[&C^{eff}_{9}(\bar{d}\gamma_{\mu}P_{L}b)\bar{\ell}\gamma_{\mu}\ell
+C_{10}(\bar{d}\gamma_{\mu}P_{L}b)\bar{\ell}\gamma_{\mu}\gamma^{5}\ell
\nonumber\\&-2\,C_{7}\bar{d}i\sigma_{\mu\nu}\frac{q^{\nu}}{q^{2}}
(m_{b}P_{R}+m_{s}P_{L})b\bar{\ell}\gamma_{\mu}\ell\,]
 , \,\,\, \label{amplitude}
\end{eqnarray}
In writing this, unitarity of the CKM matrix has been used and the
term proportional to $\lambda_t=V_{tb}^\ast V_{td}$ has been
factored out.
 where  $q$ denotes the four
momentum of the lepton pair and $C_i$'s are Wilson coefficients.
 Neglecting the terms of $O(m_q^2/m_W^2)$, $q = u, d, c$, the analytic expressions for all
Wilson coefficients, except $C_{9}^{eff}$,  can be found in
\cite{burev}. The values of $C_{7}^{eff}$ and $C_{10}$ in leading
logarithmic approximation are:
\begin{equation} \label{wilsonc7c10}
C_{7}^{eff} = -0.315,\quad  C_{10} =  -4.642
\end{equation}
only $C_9^{\rm eff}$ has weak and strong phases, i.e. :
\begin{equation}\label{c9new}
C^{eff}_{9}=\xi_{1}+\lambda_{u}\xi_{2}
\end{equation}
where the CP violating parameter $\lambda_{u}$ is as follows: \bea
  \lambda_{u}=\frac{V_{ub}^\ast V_{ud}}{V_{tb}^\ast V_{td}}=\frac{\rho (1-\rho) - \eta^2}{(1-\rho)^2 + \eta^2}
- i \frac{\eta}{(1-\rho)^2 + \eta^2}+ \cdots \quad\eea
 The
explicit expressions of functions $\xi_1$ and $\xi_2$ in $\mu=m_b$
can be found in~\cite{burev}--\cite{misiakE}:
Note that, we neglect long-distance resonant contributions  in
$C^{eff}_9$ for simplicity, a more complimentary and supplementary
analysis of the above decay has to take the long-distance
contributions, which have their origin in real intermediate
$c\bar{c}$ bound states, in addition to the short-distance
contribution into account.

The Wilson coefficients of the SM are modified by the introducing
the vector type unparticles. It is easy to see that unparticles in
this study are introduced in the way that new operators do not
appear. In other words, the full operator set for the unparticle
contributions are exactly the same as in the SM. The unparticle
effects with the SM contributions can be derived by using
$C^{\UP}_9$ and $C^{\UP}_{10}$, defined by
 \bea
 C^{\UP}_{9}(q^2) &=& C^{\rm eff}_9 + \frac{\pi}{\alpha_{em}}
\frac{C^{\ell}_R + C^{\ell}_L}{2}\tilde{\Delta}_{\UP}(q^2) e^{-i
\phi_{\UP}}\,, \nnb\\
 C^{\UP}_{10}(q^2) &=& C_{10}+ \frac{\pi}{\alpha_{em}}
\frac{C^{\ell}_R - C^{\ell}_L}{2} \tilde{\Delta}_{\UP}(q^2) e^{-i
\phi_{\UP}}\,,\nnb\\
C_7^{\UP}(q^2)&=&C_7(q^2)
 \label{eq:c7910}
 \eea
instead of $C^{\rm eff}_9$ and $C_{10}$, respectively. Where
$C_{7}$ remain the same as the SM, we can rewrite $C^{\UP}_{i}$'s
in the $m_b$ scale\cite{burev}. Then $C^{\UP}_{9}$ will be as:
\bea C^{\UP}_{9}=\xi_1^{\UP}+\lambda_u \xi_2\eea where
 \bea
 \xi^{\UP}_{1}=\xi_1+ \frac{\pi}{\alpha_{em}}
\frac{C^{\ell}_R + C^{\ell}_L}{2}\tilde{\Delta}_{\UP}(q^2) e^{-i
\phi_{\UP}}\,, \label{eq:c9up}
 \eea

Neglecting any low energy QCD corrections ($\sim 1/m_b^2$)
\cite{falk,aali} and setting the down quark mass to zero, the
unpolarized differential decay width as a function of the
invariant mass of the lepton pair is given by:
\begin{eqnarray}
\Bigg(\frac{d \Gamma}{d \hat{s}}\Bigg)_0 = \frac{G_F^2 m_b^5}{192
\pi^3} \frac{\alpha_{em}^2}{4 \pi^2} |\lambda_t|^2 (1 - \hat{s})^2
\sqrt{1 - \frac{4 \hat{m}_{\ell}^2}{\hat{s}}} \bigtriangleup^{\UP}
\label{difdecaywidth}
\end{eqnarray}
 with
\begin{eqnarray}
\bigtriangleup^{\UP}(\shat) &=& 4 \frac{(2 + \hat{s})}{\hat{s}}
\left(1 + \frac{2 \hat{m}_{\ell}^2}{\hat{s}}\right) |C^{\rm
eff}_7|^2 + (1 + 2 \hat{s}) \left(1 + \frac{2
\hat{m}_{\ell}^2}{\hat{s}}\right)
|C^{\UP}_9|^2 \nonumber \\
&& + (1 - 8 \hat{m}_{\ell}^2 + 2 \hat{s} + \frac{2
\hat{m}_{\ell}^2}{\hat{s}}) |C^{\UP}_{10}|^2  + 12 (1 + \frac{2
\hat{m}_{\ell}^2}{\hat{s}}) Re(C^{\UP*}_9 C_7^{\rm eff}) .
\label{delta}
\end{eqnarray}
The explicit expression for the unpolarized particle decay rate
$(d\Gamma/d\hat{s})_0$ has been given in (\ref{difdecaywidth}).
Obviously, it can be written as a product of a real-valued
function $r(\hat{s})$ times the function $\Delta(\hat{s})$, given
in (\ref{delta}); $(d\Gamma/d\hat{s})_0 = r(\hat{s}) \
\Delta(\hat{s})$.
In the unpolarized case, the CP-Violating asymmetry rate can be
defined by
 \begin{eqnarray}
A_{CP}^{\UP}(\hat{s})=\frac{(\frac{d\Gamma}{d\hat{s}})_0\,-\,(\frac{d\bar{\Gamma}}{d\hat{s}})_0}
{(\frac{d\Gamma}{d\hat{s}})_0\,+\,(\frac{d\bar{\Gamma}}{d\hat{s}})_0}=\frac{\Delta^{\UP}-\bar{\Delta}^{\UP}}{\Delta^{\UP}+\bar{\Delta}^{\UP}}.
\label{acps}
\end{eqnarray}
 where
\begin{eqnarray}
\frac{d\Gamma}{d\hat{s}}\equiv\frac{d\Gamma(b\rightarrow d
\ell^{+}\ell^{-})}{d\hat{s}},\quad
\frac{d\bar{\Gamma}}{d\hat{s}}\equiv\frac{d\bar{\Gamma}(\bar{b}\rightarrow
\bar{d} \ell^{+}\ell^{-})}{d\hat{s}} \label{gama0}
\end{eqnarray}
where, $(d\bar{\Gamma}/d\hat{s})_0$ can be obtained from
$(d\Gamma/d\hat{s})_0$ by making the replacement \bea
\label{e5714} C_9^{\UP} = \xi_1^{\UP}+\lambda_u\xi_2 \rar
\bar{C}_9^{\UP}= \xi_1^{\UP}+\lambda_u^\ast\xi_2~. \eea Note that
the  term proportional to $\lambda_u$, CP violating parameter
remains the same as the SM. Moreover, the CP violating parameter
just enters into the $C_9^{\UP}$ expression same as the SM ones.
Consequently, the rate for anti-particle decay can be obtained by
the following replacement in the Eq. (\ref{delta}): \bea
\bar{\Delta}^{\UP}=\Delta^{\UP}_{|\lambda_u\rightarrow
\lambda_u^\ast}\eea Using (\ref{acps}), the CP violating asymmetry
is evaluated to be:
\begin{equation}\label{asymcomp}
A_{CP}^{\UP}(\hat{s}) = \frac{ -2\,  {\rm
Im}(\lambda_{u})\Sigma^{\UP}(\hat{s})}{\Delta^{\UP}(\hat{s}) + 2\,
{\rm Im}(\lambda_{u})\Sigma^{\UP}(\hat{s})}\approx -2\,  {\rm
Im}(\lambda_{u})\frac{\Sigma^{\UP}(\hat{s})}{\Delta^{\UP}(\hat{s})}.
\end{equation}
In (\ref{asymcomp}),
\begin{eqnarray}\label{sigma}
\Sigma^{\UP}(\shat)&=& {\rm Im} [\xi_1^{\UP\ast}
\,\xi_2]f_+(\hat{s}) + {\rm Im} (C_7^{eff\ast} \xi_2^{})
f_1(\hat{s})\nonumber\\
f_+(\hat{s})&=&(1 + 2 \hat{s}) \left(1 + \frac{2\hat{m}_{\ell}^2}{\hat{s}}\right)\nonumber\\
f_1(\hat{s})&=& 12 (1 + \frac{2 \hat{m}_{\ell}^2}{\hat{s}})
\end{eqnarray}
Before turning to a derivation of CP violating asymmetries for the
case of polarized final state leptons, it is necessary  to remind
the calculation of the lepton polarization.  The spin direction of
a lepton can be described by setting  a reference frame with three
orthogonal unit vectors $S_L$, $S_N$ and $S_T$, such that
\begin{eqnarray}\label{eq:unitvecdef}
S_L &=& \frac{\plep}{|\plep |}~,
\nonumber\\
S_N &=&\frac{p_d \times \plep}{|p_d \times \plep |}~,
\\
S_T &=& S_N \times S_L ~, \nonumber
\end{eqnarray}
where $p_d$ and $\plep$ are the three momentum vectors of the d
quark and the $\ell^-$ lepton, respectively, in the $\ell^+
\ell^-$ center-of-mass(CM) system.
For a given lepton $\ell^-$ spin direction $\vec{n}$, which is a
unit vector in the $\ell^-$ rest frame, the differential decay
spectrum is of the form \cite{Kruger:1996cv}
\begin{equation}
\frac{d\Gamma(\shat, \vec{n})}{d\shat} = \frac{1}{2}
\left(\frac{d\Gamma(\shat)}{d\shat}\right)_0 \Big[ 1 + (P_L e_L +
P_T e_T + P_N e_N) \cdot \vec{n} \Big], \label{eq:poldecay}
\end{equation}
where the polarization components $P_i$ (${\rm i = L,N,T}$) are
obtained from the relation
\begin{equation}
P_i(\shat) =  \frac{d \Gamma (\vec{n} = e_i)/d \shat  -  d \Gamma
(\vec{n} = -e_i)/d\shat} {d \Gamma (\vec{n} = e_i)/d\shat +  d
\Gamma (\vec{n} = -e_i)/d\shat
}=\frac{\Delta^{\UP}_i(\shat)}{\Delta^{\UP}(\shat)}~ .
\label{leppolldef}
\end{equation}
The three different  polarization asymmetries are:
\begin{eqnarray}\label{pol}
P_L(\shat)&=&\frac{\Delta^{\UP}_L(\shat)}{\Delta^{\UP}(\shat)}=
\frac{v}{\Delta^{\UP}(\shat)} \Big[12{\rm Re}({C_7^{eff}}
C_{10}^{\UP\ast}) +  2{\rm Re}({C_9^{\UP}}C_{10}^{\UP\ast})(1 ~ +
2\shat)\Big],
\nonumber \\
P_T(\shat)  &=&
\frac{\Delta^{\UP}_T(\shat)}{\Delta^{\UP}(\shat)}=\frac{3 \pi
\mlhat}{2 \Delta^{\UP} (\shat) \sqrt{\shat}}
 \Big[2{\rm Re}({C_7^{eff}} C_{10}^{\UP\ast})
 -4 {\rm Re}(C_7^{eff} C_9^{\UP\ast})  \nonumber \\ &&-
\frac{4}{\shat}|{C_7^{eff}}|^2 \phantom{=}+ {\rm
Re}({C_9^{\UP}}C_{10}^{\UP\ast})- |{C_9^{\UP}}|^2 \shat\Big],
\nonumber\\
P_N(\shat)  &=& \frac{\Delta^{\UP}_N(\shat)}{\Delta^{\UP}(\shat)}=
\frac{3 \pi \mlhat v}{2 \Delta^{\UP} (\shat)} \sqrt{\shat}\ {\rm
Im}({C_9^{\UP}}^*C_{10}^{\UP})~, \label{leppollasym}
\end{eqnarray}
The study of the above-mentioned asymmetries is interesting in
probing new physics. As it is obvious that any alteration in the
Wilson coefficients leads to changes in the polarization
asymmetries.

Now, we define the polarized CP asymmetry which is:
\begin{equation}
A_{CP}(\shat) = \frac{\frac{d\Gamma(\shat,\vec{n})}{d \shat} -
\frac{d\bar\Gamma(\shat,\bar{\vec{n}})}{d \shat}}
{(\frac{d\Gamma(\shat)}{d \shat})_0 + (\frac{d\bar\Gamma(\shat)}{d
\shat })_0}~, \label{eq:acpn}
\end{equation}
where
\begin{eqnarray}
\frac{d\Gamma(\shat,\vec{n})}{d \shat} &=& \frac{d\Gamma(b \to d
\ell^+ \ell^-(\vec{n}))}{d \shat }\ ,
\nonumber\\
\frac{d\bar\Gamma(\shat,\vec{\bar n})}{d \shat } &=&
\frac{d\Gamma(\bar b \to \bar d \ell^+(\vec{\bar n}) \ell^-)}{d
\shat }~,
\end{eqnarray}
here, $\vec{n}$ and $\vec{\bar n}$ are  the spin directions for
$\ell^-$ and $\ell^+$ for $b$-decay and $\bar b$-decay,
respectively, and $i=L, ~N,~T$. Taking into account the fact that
$\vec{\bar e}_{L,N}= -\vec{e}_{L,N}$, and $\vec{\bar e}_{T}=
\vec{e}_{T}$, we obtain \bea \label{e5718} A_{CP}(\vec{n}=\pm
\vec{e}_i) \es \frac{1}{2} \left\{
\frac{\ds{(\frac{d\Gamma}{d\hat{s}})_0} -
 \ds{( \frac{d\bar{\Gamma}}{d\hat{s}})_0}}
{\ds{( \frac{d\Gamma}{d\hat{s}})_0} +
 \ds{( \frac{d\bar{\Gamma}}{d\hat{s}})_0}}
\pm \frac{\ds{( \frac{d\Gamma}{d\hat{s}})_0 P_i} - \ds{
\Bigg((\frac{d\Gamma}{d\hat{s}})_0
P_i\Bigg)_{|\lambda_u\rightarrow \lambda_u^\ast}}} {\ds{(
\frac{d\Gamma}{d\hat{s}})_0} +
 \ds{( \frac{d\bar{\Gamma}}{d\hat{s}})_0}}
\right\}~. \eea Using Eq. (\ref{pol}), we get from Eq.
(\ref{e5718}), \bea \label{e5719} A_{CP} (\vec{n}
=\pm\vec{e}_i)&\approx& \frac{1}{2} \left\{
\frac{\Delta^{\UP}-\bar{\Delta}^{\UP}}
{\Delta^{\UP}+\bar{\Delta}^{\UP}} \pm
\frac{\Delta^{\UP}_i-\bar{\Delta}^{\UP}_i}{\Delta^{\UP}+\bar{\Delta}^{\UP}}\right\}~, \nnb \\
\es \frac{1}{2} \left\{A_{CP} (\hat{s}) \pm  A_{CP}^i (\hat{s})
\right\}~, \eea where the upper sign in the definition of $\delta
A_{CP}$ corresponds to $L$ and $N$ polarizations, while the lower
sign corresponds to $T$ polarization.

The $ A_{CP}^i (\hat{s})$ terms in Eq. (\ref{e5719}) describe the
modification to the unpolarized decay width, which can be written
as: \bea \label{e5720}  A_{CP}^i (\hat{s}) \es \frac {-4 \mbox{\rm
Im}( \lambda_u )
\Sigma^i(\hat{s})}{\Delta^{\UP}(\hat{s})+\bar{\Delta}^{\UP}(\hat{s})}~, \nnb \\
\ap -2 \mbox{\rm Im} (\lambda_u) \frac{
\Sigma^i(\hat{s})}{\Delta^{\UP}(\hat{s})}~. \eea where, the
explicit expressions for $\Sigma^i(\hat{s}),~(i=L,N,T)$, are as
follows: \bea\label{acppol}
 \Sigma^L(\shat)&=& v {\rm Im}(C_{10}^{\UP\ast} \xi_2)(1+2\shat)\nnb\\
  \Sigma^T(\shat)&=& \frac{3 \pi \hat{m}_\ell}{2\sqrt{\shat}}
  \Bigg[2{\rm Im}(C_{7}^{\rm eff} \xi_2^{\ast})+\frac{1}{2}{\rm Im}(C_{10}^{\UP\ast} \xi_2)-\shat{\rm Im}(\xi_{1}^{\UP\ast} \xi_2)\Bigg]\nnb\\
  \Sigma^N(\shat)&=& \frac{3 \pi \hat{m}_\ell}{2\sqrt{\shat}}v \Bigg[\frac{\shat}{2}{\rm Re}(C_{10}^{\UP\ast} \xi_2)\Bigg]\eea
It is interesting to note that the polarized CP asymmetries have
different combinations involving the imaginary and real parts of
the $C_{10}^{\UP}$ which doesn't appear in unpolarized CP
asymmetry. The study of the polarized CP asymmetry beside the
unpolarized CP asymmetry with unparticle contributions will give
us more information about the unparticle parameters. In
particular, when $C_{L}^\ell=-C_{R}^\ell$ in (\ref{eq:c7910}), the
unparticle contribution vanishes in the $C_{9}^{\UP}$. In such
situation, the unparticle effects in CP asymmetry just appear in
the polarized CP asymmetries.

\section{Numerical Analysis and Discussion}
We try to analyze the dependency of the unpolarized and polarized
direct CP asymmetries on the unparticle parameters .
 We will use the next--to--leading order
logarithmic approximation for the SM  values of the Wilson
coefficients $C_9^{eff},~C_7^{\rm eff}$ and $C_{10}^{\rm eff}$
Ref.~\cite{R46215,R46216} at the scale $\mu=m_b$. It is worth to
mention that, beside the short distance contribution, $C_9^{eff}$
has also long distance contributions resulting from  the real
$\bar c c$ resonant states of the $J/\psi$ family. In the present
study, we do not take the long distance effects into account.
Furthermore, one finds that significant contributions of
unparticle occurs at small region of $\shat$ which is free of long
distance effects(obviously, the unparticle contributions for $\mu$
channel is significant than the $\tau$ channel since small $\shat$
region($\shat\sim 0.0$) for $\tau$ channel is absent by
kinematical consideration). One can confirm the above statement by
looking at Eqs.~(\ref{eq:lang_UP}) and (\ref{eq:uprop}) where at
the small $\shat=q^2/m_b^2$ region the dependency of the
propagator is as follows: \bea
 \left[\frac{1}{q^2}\left(\frac{q^2}{\Lambda^2_{\UP}}\right)^{d_{\UP}-1}\right]^{2}\,.
  \eea

The SM parameters we used in this analysis are shown in Table 1:
\begin{table}[h]
        \begin{center}
        \begin{tabular}{|l|l|}
        \hline
        \multicolumn{1}{|c|}{Parameter} & \multicolumn{1}{|c|}{Value}     \\
        \hline \hline
        $\alpha_{em}$                   & $1/129$ (GeV) \\
        $m_{u}$                   & $2.3$ (MeV) \\
        $m_{d}$                   & $4.6$ (MeV) \\
        $m_{c}$                   & $1.25$ (GeV) \\
        $m_{b}$                   & $4.8$ (GeV) \\
        $m_{\mu}$                   & $0.106$ (GeV) \\
        $m_{\tau}$                  & $1.780$ (GeV) \\
        \hline
        \end{tabular}
        \end{center}
\caption{The values of the input parameters used in the numerical
          calculations.}
\label{input}
\end{table}

The allowed range for the Wolfenstein parameters are: $0.19 \le
\rho \le 0.268$ and $0.305 \le \eta \le 0.411$ \cite{R5745} where,
in the present analysis they are set as $\rho=0.25$ and
$\eta=0.34$.

The direct CP asymmetries  depend on both $\shat$ and the new
parameters coming from unparticle stuff. We eliminate the variable
$\hat{s}$ by performing integration over $\shat$ in the allowed
kinematical region. The averaged direct CP asymmetries are defined
as: \bea {\cal B}_r&=&\ds \int_{4
m_\ell^2/m_{b}^2}^{(1-\sqrt{\hat{r}_{d}})^2}
 \frac{d{\cal B}}{d\hat{s}}
 d\hat{s},\,\,~~(\hat{r}_{d}=\frac{m_d^2}{m_b^2})
\nnb\\\lla A^{i}_{CP} \rra &=& \frac{\ds \int_{4
m_\ell^2/m_{b}^2}^{(1-\sqrt{\hat{r}_{d}})^2} A^{i}_{CP}
\frac{d{\cal B}}{d\hat{s}} d\hat{s}} {{\cal{B}}_r}~. \eea
 At this
stage, we discuss our restrictions for  free parameters coming out
of the unparticle:
\begin{itemize}
\item It is important to note that while the discontinuity across
the cut is not singular for integer $d_{\UP}>1$, the propagator
(Eq. \ref{propage}) is singular because of the $sin(d_{\UP} \pi)$
in the denominator. Some researchers  believe that this is a real
effect\cite{Georgi2}. These integer values describe multiparticle
cuts and the mathematics tells us that we should not  try to
describe them with a single unparticle field.

Moreover, the lower bounds for the scaling dimensions of the gauge
invariant vector operators of a CFT are $d_{\UP}\geq 2$ and
$d_{\UP}\geq 3$\cite{Grinstein} for non-primary and primary vector
operators, respectively. We obtain that for $d_{\UP}> 2$ the
unparticle effects on physical observables(branching ratio, CP
asymmetry and so on) almost vanish because $\tilde{\Delta}(p^2)$
is negligible for $p<\Lambda_{\UP}$(see Eq.~\ref{propage2}).

 We focus on  $1 < d_{\UP} < 2$, the bound that is allowed for
 transverse $O^{\mu}_{\UP}$ or for non-gauge invariant vector
 operators of the CFT. Also, it is consistent
with the $b\to s\ell^+\ell^-$ rate\cite{Chen5} and $B_s$
mixing\cite{Chen6}. We also assume that the virtual effect of
unparticles are gentlest away from the integer boundaries. On the
other hand, the momentum integrals converges for $d_{\UP}<2$
\cite{Liao1}.
\item $C^{t}_{L}$ is always associated with $C^{\ell}_R$  and
$C^{\ell}_{R}$(see Eq.~(\ref{eq:c7910}). For simplicity, we set
$C^{\ell}_R= C^{\ell}_L$ or $C^{\ell}_R= -C^{\ell}_L$. We will set
new parameters to be $C^t_L C^{\ell}_L=C^t_L
C^{\ell}_R=\lambda^{\ell}_V$ and $C^t_L C^{\ell}_L=-C^t_L
C^{\ell}_R=\lambda^{\ell}_A$ and choose the
 $\lambda^{\ell}_{V[A]}=0.005$, $0.01$ and $0.05$ which is consistent with the $b\to
s\ell^+\ell^-$ rate\cite{Chen5}.
 \item  We take the energy
scale  $\Lambda_{\UP}=1\,(TeV)$ and study $d_{\UP}$ dependence of
the polarized and unpolarized CP asymmetry.
\end{itemize}
CP asymmetry is good candidate(unlike the other physical
observables i.e,. branching ratio, forward-backward asymmetry and
...) to probe the unique unparticle phase. The other physical
observables  can utilize to give strong constraints on the
unparticle parameters except the phase, {\it i.e.} on the
unparticle couplings to leptons such as
$\lambda^{\ell}_{V(A)}\sim\{0.005-0.05\}$\cite{Chen5}. Moreover,
our numerical analysis confirm the result  of the \cite{Chen5}
where the  branching ratio(BR) of $b\to s(d) \ell^{+} \ell^{-}$
decay depict the strong enhancement at the low value of the scale
dimension $d_{\UP}\sim 1.1$ with respect the SM value. As a
natural consequence of this feature, the averaged value of
asymmetries will vanish unless they depict stronger enhancement
than the BR.

 The contributions of unparticle to  the CPA of $b\to d \ell^{+}
\ell^{-}$ in terms of the  values for the common parameters  are
presented in Fig.~3--10. The horizontal thin lines are the SM
contributions, the dashed lines and dash--pointed lines correspond
to the different $\lambda^{\ell}_{A[V]}=0.005$, $0.01$ and $0.05$
, respectively. From theses figures, we conclude that:
\begin{itemize}
\item $\langle A_{CP} \rangle$ for both $\mu$ and $\tau$ leptons
depicts strong dependency on the unparticle effects(for $\mu$ case
the dependency is stronger that $\tau$  case as we discussed
above). While it is suppressed to the zero value by the unparticle
contributions at lower values of the scale dimension $d_{\UP}\sim
1.1$. Its value is close to  the SM value at the higher values of
the scale dimension $d_{\UP}\sim 1.9$. Moreover, the sensitivity
for different values of the $\lambda_A$ is stronger and more
interesting than the $\lambda_V$ values.  While for different
$\lambda_V$ values, $\langle A_{CP} \rangle$ is just decreasing in
terms of the $d_{\UP}$, but for $\lambda_A$ it is increasing,
decreasing and changing the sign(see figs. 3, and 4).

\item $\langle A_{CP}^L \rangle$ for both $\mu$ and $\tau$ leptons
shows strong dependency on the unparticle parameters. While it is
suppressed to the zero value by the unparticle contributions at
lower values of the scale dimension $d_{\UP}\sim 1.1$(see figs. 5,
and 6), its value is close to  the SM value at the higher values
of the scale dimension $d_{\UP}\sim 1.9$. The situation for the
$\mu$ leptons is  much more interesting. While the SM value is
about few percent, but it receive sizable and measurable
contribution up to $10\%$ from unparticle effects(see figs. 5). As
$\langle A_{CP}^L \rangle$ and $\langle A_{CP} \rangle$ are
sensitives  to the $C^{\UP}_{10}$ and $C^{\UP}_9$, respectively,
thus, the study of $\langle A_{CP}^L \rangle$ beside $\langle
A_{CP} \rangle$ is supplementary and complementary to study
unparticle effects. More precisely, unlike $\langle A_{CP}
\rangle$, $\langle A_{CP}^L \rangle$ shows stronger dependency on
the the different values of the $\lambda_A$ than the $\lambda_V$
values.

\item $\langle A_{CP}^T \rangle$ is generally sensitive to the
unparticle contributions for both $\mu$ and $\tau$ channels.
While, the SM values of $\langle A_{CP}^T \rangle$ almost
vanishes,  the unparticle contributions lead to  sizable deviation
from the SM values,(see figs. 7 and 8). This sizable discrepancy
from the SM values can be measured in future experiments like LHC
and ILC.

\item Either the SM value or its value with unparticle
contributions for $\langle A_{CP}^N \rangle$ is negligible(see
figs. 9 and 10).
\end{itemize}
At the end, the quantitative estimation about the accessibility to
measure the various physical observables are in order. An
observation of a 3$\sigma$ signal for CP asymmetry of the order of
the $1\%$ requires about $\sim 10^{10}$ $B \bar{B}$
pair\cite{Kruger:1996cv}. For $b \to d \ell^+ \ell^-$ measurement
a good $d$-quark tagging is necessary to distinguish it from much
more stronger $b \to s \ell^+ \ell^-$ decay signal. Putting aside
this challenging task, the number of $B \bar{B}$ pairs, expected
to produce at  LHC, are about $\sim 10^{12}$. As a result of
comparison of these values, we conclude that a typical asymmetry
of (${\cal A}=1\%$) is certainly detectable at LHC.

In conclusion, firstly, we obtain that the unparticle effects on
physical observables i.e., branching ratio and CP asymmetry for $b
\to d(s) \ell^+ \ell^-$ decays when $d_{\UP}\geq 2$ vanish.
Secondly, for $1<d_{\UP}< 2$, the CP asymmetry for polarized and
unpolarized lepton cases are studied within the unparticle
contributions in the CPA of the $b \to d \ell^+ \ell^-$ decays. We
obtain that the unpolarized and polarized CP asymmetries  are
strongly sensitive to the unparticle effects. In particular, the
CPA  for small values of scale dimension $d_{\UP} \sim 1.1$
suppresses to zero and for its definite values the CPA enhances
considerably and changes its sign with respect to the
corresponding SM value. The other parameters of the scenario
studied are the $\UP$-fermion-fermion couplings, the energy scale
and the dependencies of the CPA  to these free parameters are also
strong. We show that a measurement of the magnitude and sign of
the unpolarized and polarized asymmetries can be instructive in
order to test the possible signals coming from the unparticle
physics.
\section{Acknowledgment}
The authors would like to thank T. M. Aliev for his useful
discussions.
\newpage
\section*{Figure captions}

{\bf Fig. (3)} The dependence of the $\la {\cal A}_{CP}\ra$ for
the $b \rar d \mu^+ \mu^-$ decay on $d_{\UP}$ for three different
values of $\lambda_V:0.005,~0.01,~0.05$ and
$\lambda_A:0.005,~0.01,~0.05$ in the fixed value of
$\Lambda_{\UP}=1$TeV.\\  {\bf Fig. (4)} The same as in Fig. (1),
but for the $\tau$ lepton.\\  {\bf Fig. (5)} The dependence of the
$\la {\cal A}_{CP}^L\ra$ for the $b \rar d \mu^+ \mu^-$ decay on
$d_{\UP}$ for three different values of
$\lambda_V:0.005,~0.01,~0.05$ and $\lambda_A:0.005,~0.01,~0.05$
 in the fixed value of $\Lambda_{\UP}=1$TeV.\\
{\bf Fig. (6)} The same as in Fig. (3), but for the $\tau$
lepton.\\ {\bf Fig. (7)} The dependence of the $\la {\cal
A}_{CP}^T\ra$ for the $b \rar d \mu^+ \mu^-$ decay on $d_{\UP}$
for three different values of $\lambda_V:0.005,~0.01,~0.05$ and
$\lambda_A:0.005,~0.01,~0.05$
 in the fixed value of $\Lambda_{\UP}=1$TeV.\\
{\bf Fig. (8)} The same as in Fig. (5), but for the $\tau$ lepton
.\\ {\bf Fig. (9)}  The dependence of the $\la {\cal A}_{CP}^N\ra$
for the $b \rar d \mu^+ \mu^-$ decay on $d_{\UP}$ for three
different values of $\lambda_V:0.005,~0.01,~0.05$ and
$\lambda_A:0.005,~0.01,~0.05$
 in the fixed value of $\Lambda_{\UP}=1$TeV.\\
{\bf Fig. (10)} The same as in Fig. (5), but for the $\tau$
lepton.\\

\newpage

\begin{figure}
\vskip 1.5 cm
    \includegraphics{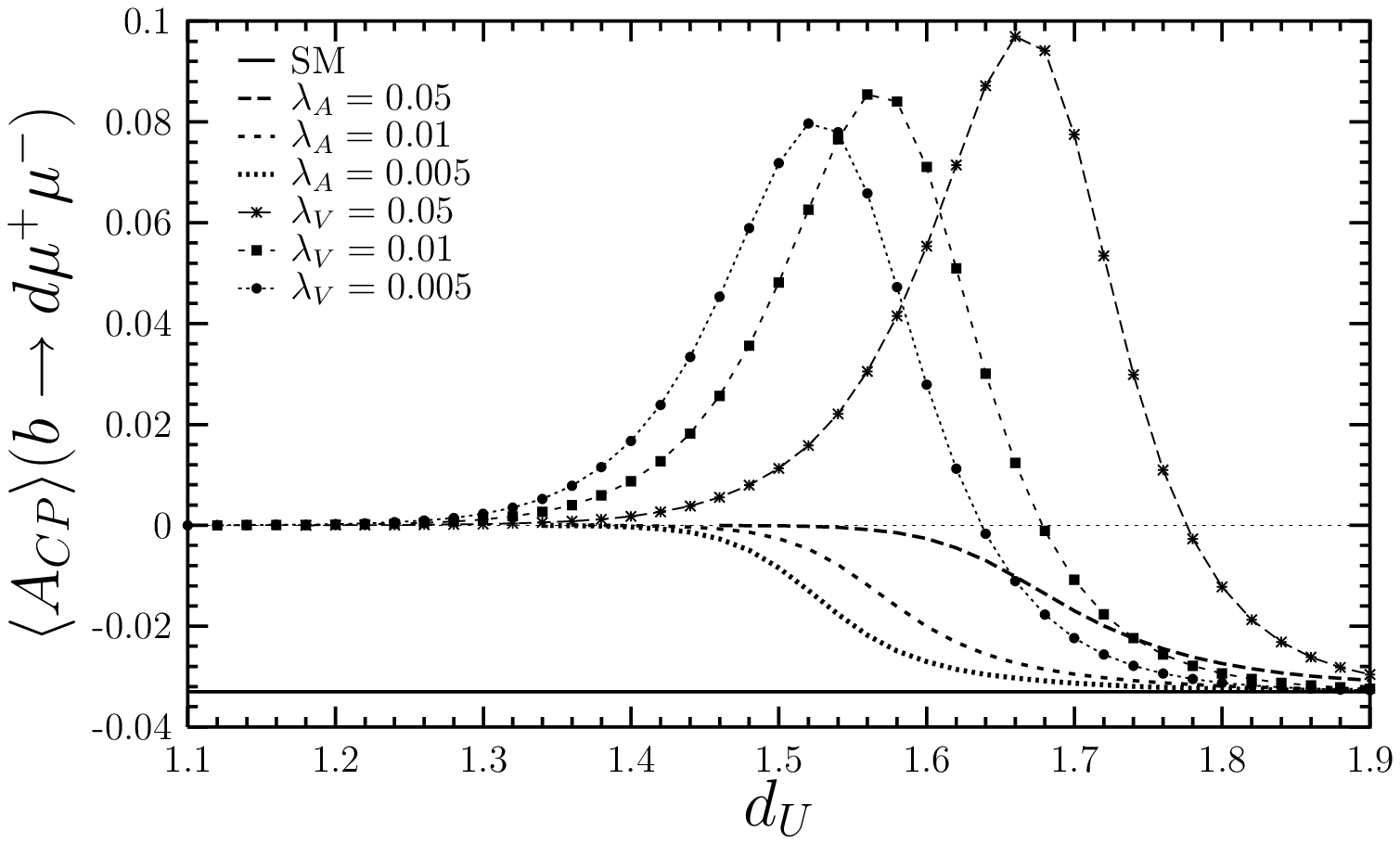}
\vskip 7.8 cm \caption{}
\end{figure}
\begin{figure}
\vskip 1.5 cm
    \includegraphics{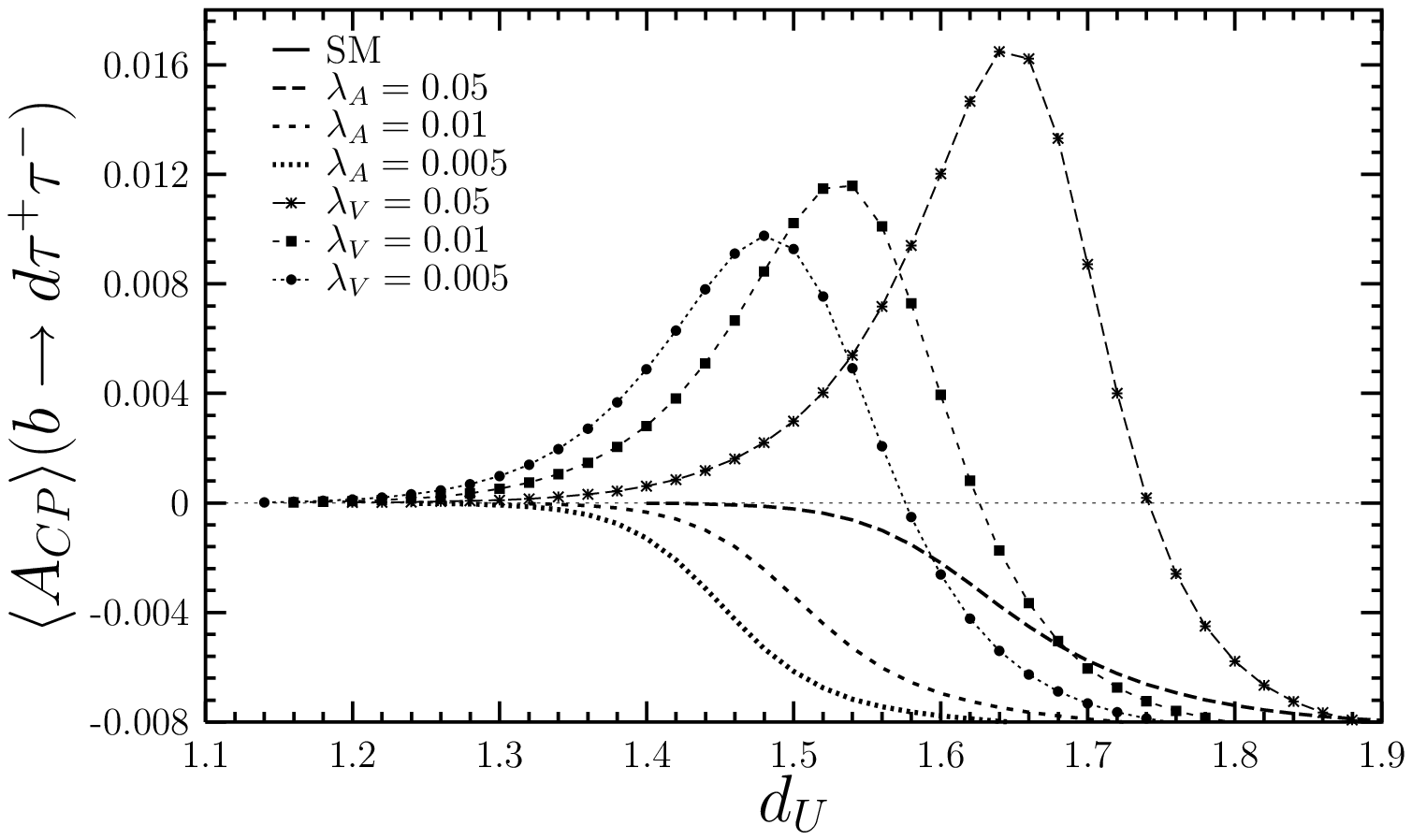}
\vskip 7.8cm \caption{}
\end{figure}

\begin{figure}
\vskip 2.5 cm
    \includegraphics{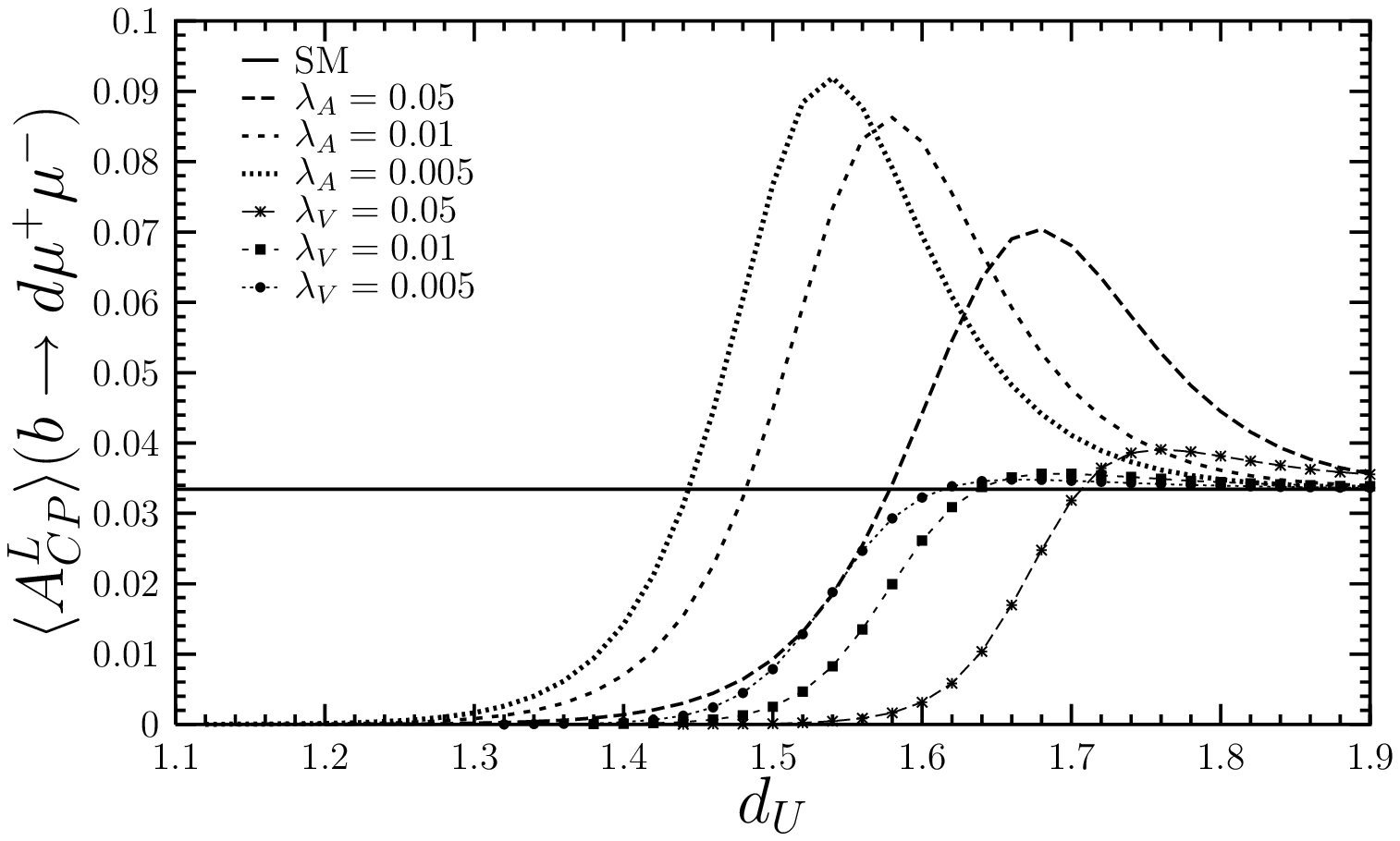}
\vskip 7.8 cm \caption{}
\end{figure}

\begin{figure}
\vskip 1.5 cm
    \includegraphics{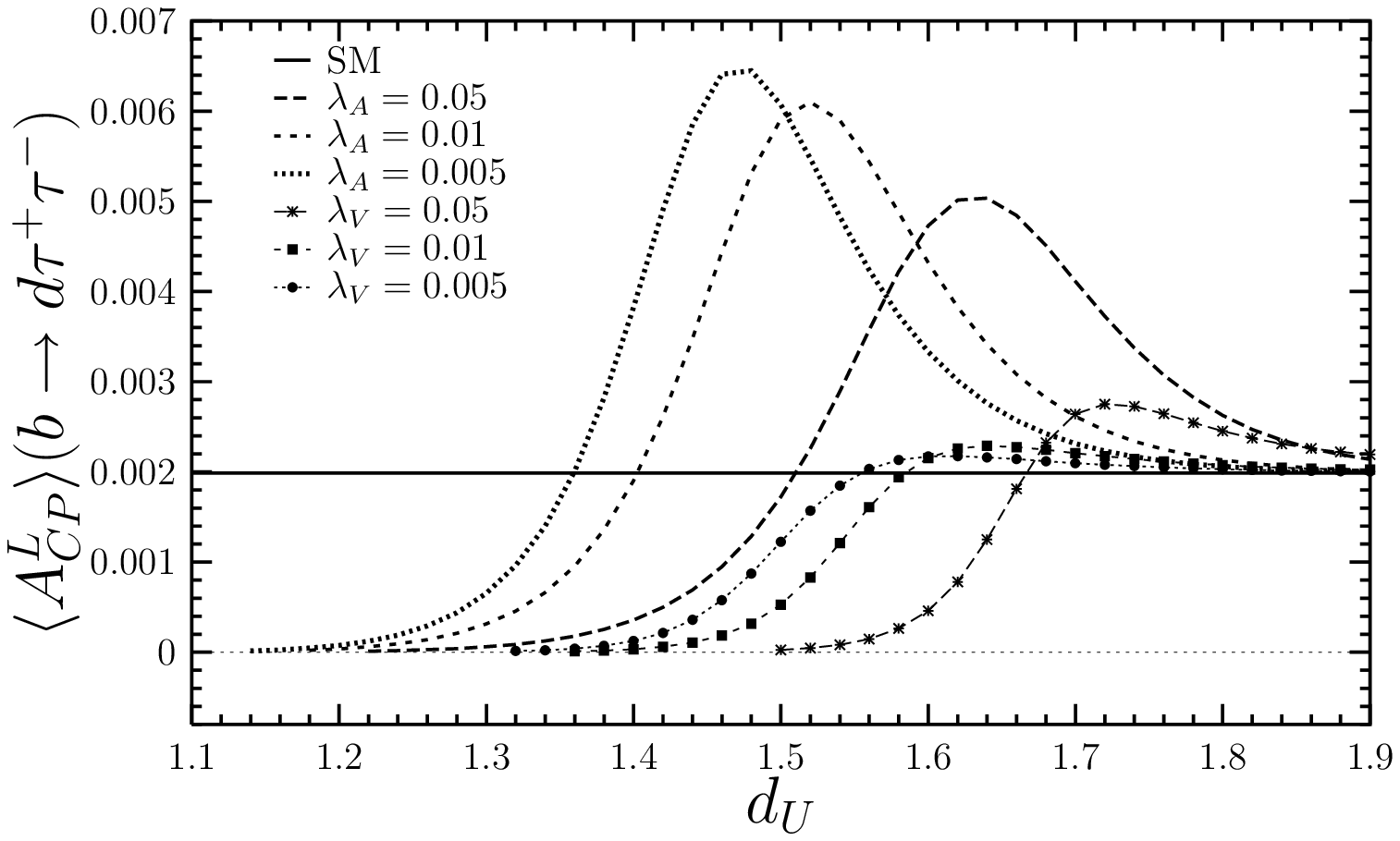}
\vskip 7.8cm \caption{}
\end{figure}

\begin{figure}
\vskip 2.5 cm
    \includegraphics{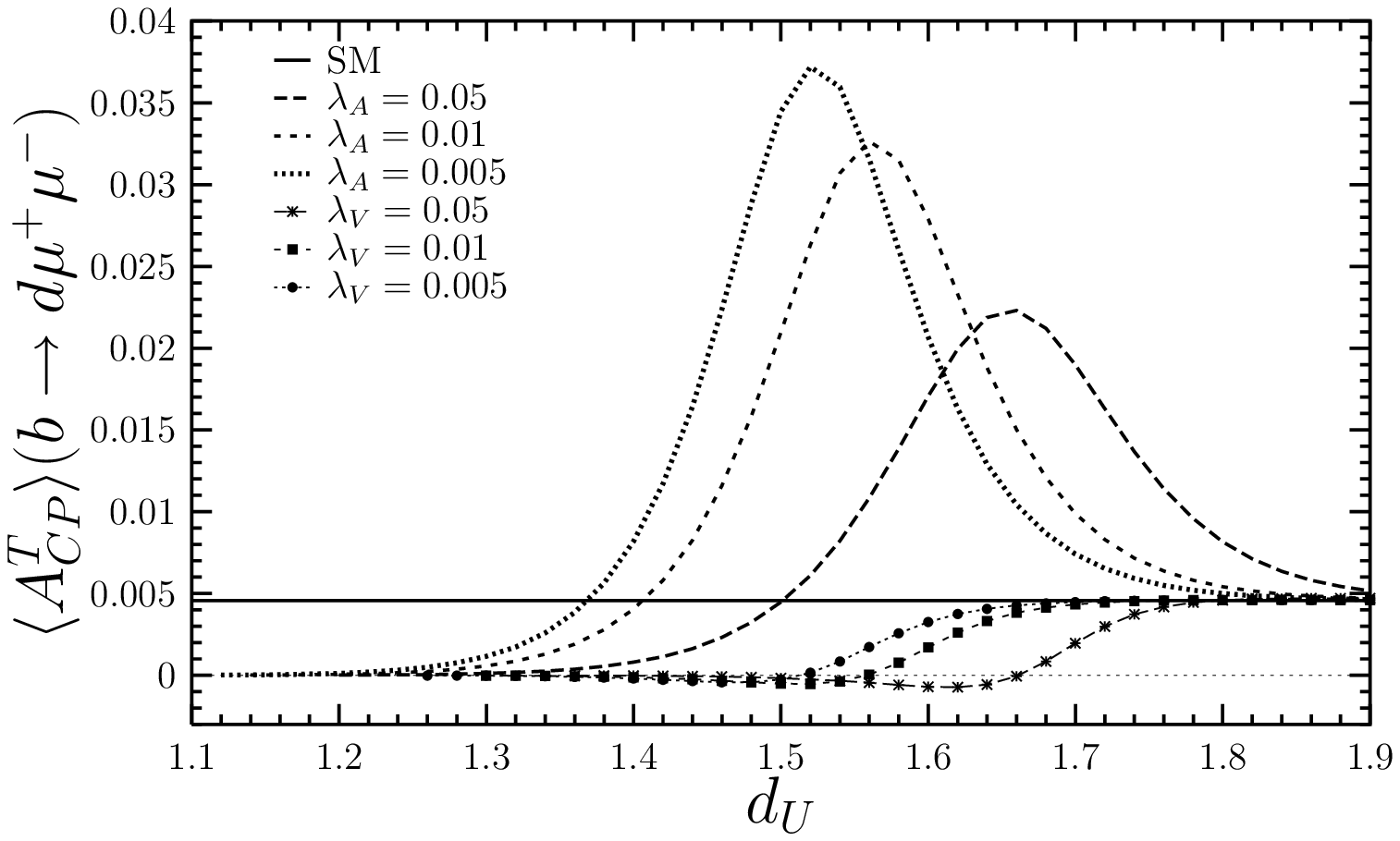}
\vskip 7.8 cm \caption{}
\end{figure}

\begin{figure}
\vskip 2.5 cm
    \includegraphics{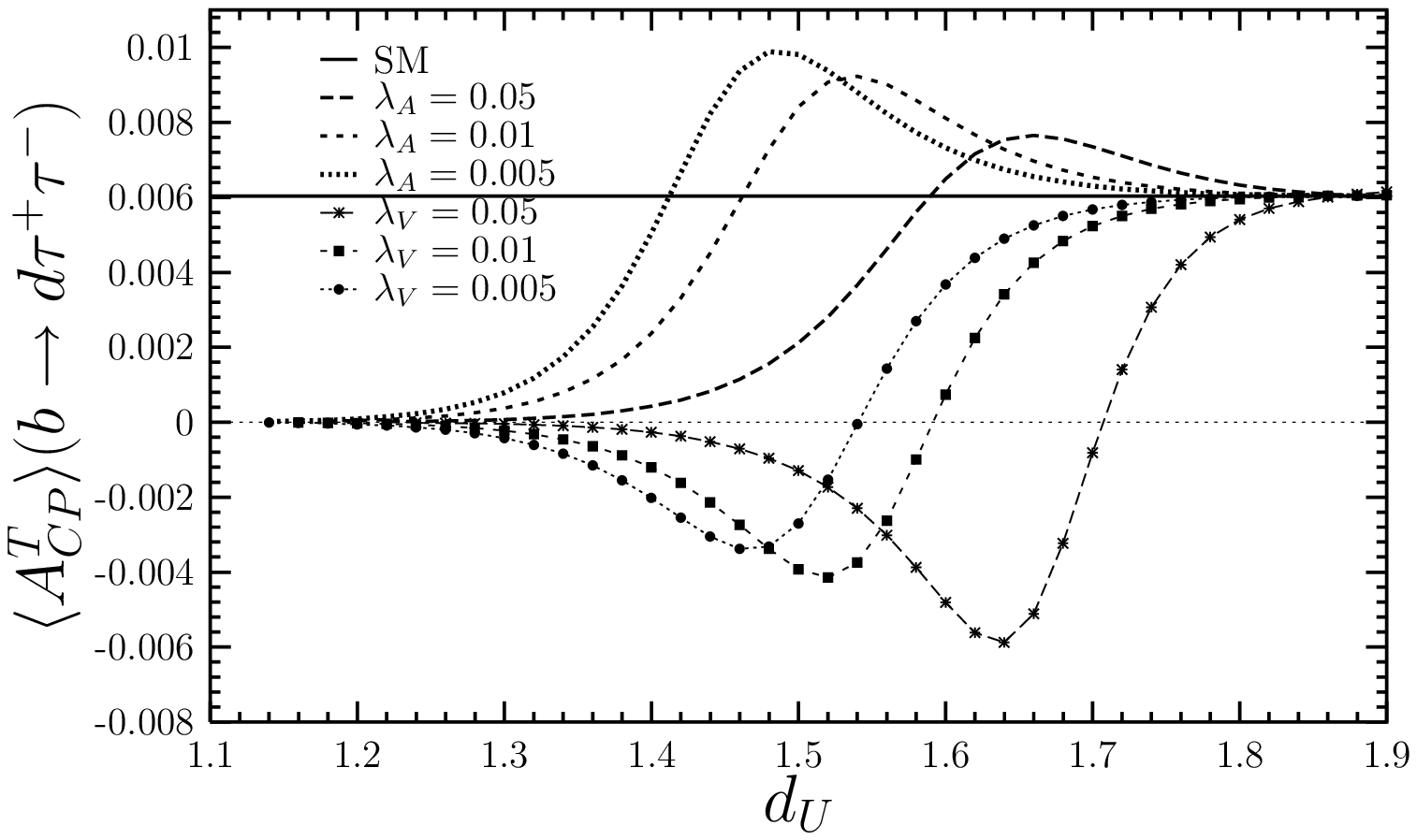}
\vskip 7.8 cm \caption{}
\end{figure}

\begin{figure}
\vskip 2.5 cm
    \includegraphics{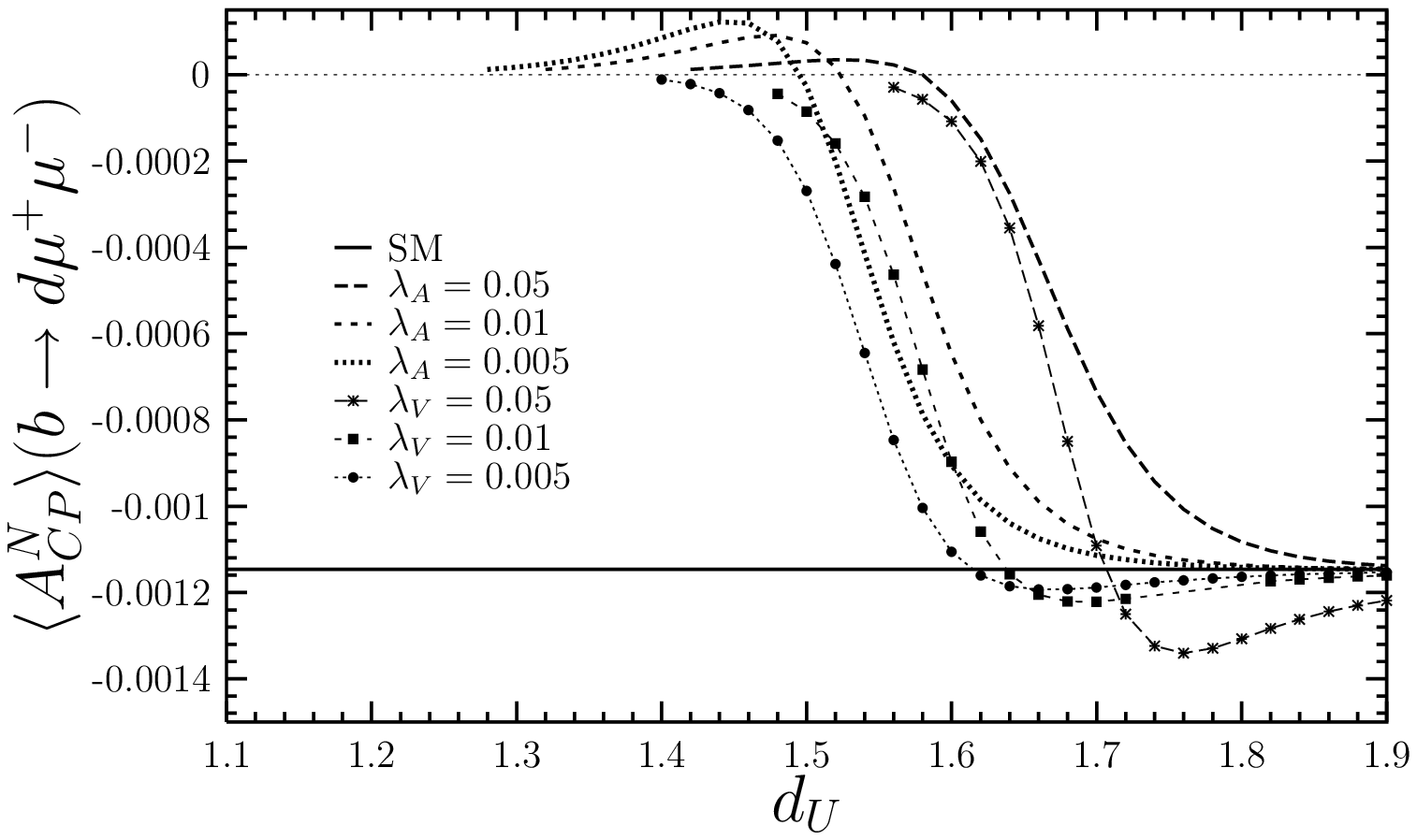}
\vskip 7.8 cm \caption{}
\end{figure}

\begin{figure}
\vskip 2.5 cm
    \includegraphics{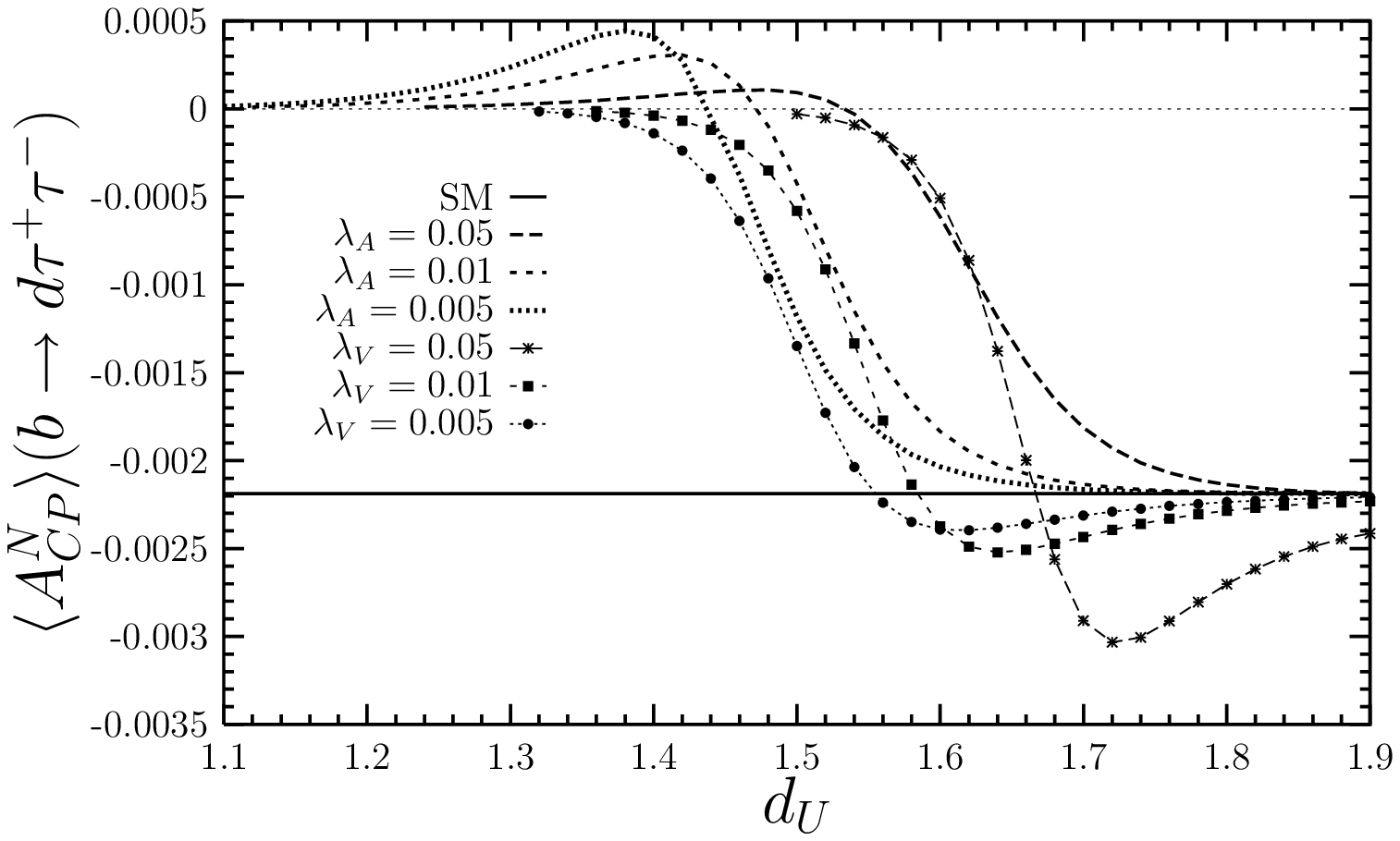}
\vskip 7.8 cm \caption{}
\end{figure}

\end{document}